\documentclass[epj]{svjour}
\usepackage{latexsym}
\usepackage{graphicx}
\usepackage{amssymb}
\begin{document}
\title{Compact $U(1)$ gauge theories in $2+1$ dimensions 
and the physics of low dimensional insulating materials}
\author{F. S. Nogueira\inst{1} \and J. Smiseth\inst{2} 
\and E. Sm{\o}rgrav\inst{2} \and A. Sudb{\o}\inst{2}
}                     
%
%
\institute{Institut f{\"u}r Theoretische Physik, 
Freie Universit{\"a}t Berlin,
D-14195 Berlin, Germany \and Department of Physics, Norwegian University of
Science and Technology, N-7491 Trondheim, Norway}
\date{Received: date / Revised version: date}
%
\abstract{
Compact abelian gauge theories in $d=2+1$ dimensions arise 
often as an effective field-theoretic description of 
models of quantum insulators. In this paper we review 
some recent results about the compact abelian Higgs 
model in $d=2+1$ in that context.  
} 
\authorrunning{F.S. Nogueira {\it et al.}}
\titlerunning{Compact $U(1)$ gauge theories in $2+1$ dimensions}
\maketitle
\section{Introduction}
\label{intro}

Effective abelian gauge theories of doped Mott insulators are very 
popular in the condensed matter literature. Despite the enormous 
amount of papers published over the last ten years, there are still 
many points of disagreement among researchers, mostly  related to poorly 
understood issues  \cite{Kim,Nayak,Nagaosa,Matsui}. One key point is the 
interplay between the confinement/deconfinement properties of the many 
theories available, and spin-charge separation. It is usually believed 
that if deconfinement occurs, then spin-charge separation would also 
occur. However, it remains a  controversial issue if deconfinement is 
possible for gauge fields coupled to matter with the fundamental charge. 
The problem here is that most $U(1)$ effective gauge theories are 
studied in $d=2+1$ space-time dimensions. For this dimensionality 
$U(1)$ gauge fields are strongly confining. For instance, if 
we neglect the coupling to matter fields we obtain that 
pure {\it compact} Maxwell theory permanently confines electric charges 
in $d=2+1$ dimensions \cite{Polyakov}. This is in contrast 
with pure compact Maxwell theory in $d=3+1$ dimensions where 
a deconfinement transition occurs \cite{Peskin}. A natural question 
to ask in this context is whether the coupling to matter fields is able 
to change the permanent confinement in $2+1$ dimensions. One example 
where the answer to this question is affirmative corresponds to the case where 
bosons are coupled to an abelian gauge field with {\it integer} charge 
$q>1$ \cite{FradShe}. This is most easily understood by considering the 
lattice gauge theory version of such a model, whose action is given by

\begin{equation}
\label{AHM}
S=-\beta\sum_{i,\mu}[\cos(\nabla_\mu\theta_i-qA_{i\mu})-1]
-\kappa\sum_{i,\mu,\nu}[\cos(F_{i\mu\nu})-1],
\end{equation}
where we have $\nabla_\mu\theta_i\equiv\theta_{i+\hat{\mu}}-\theta_i$ and 
$F_{i\mu\nu}\equiv\nabla_\mu A_{i\nu}-\nabla_\nu A_{i\mu}$. 
The above action corresponds to the compact lattice abelian Higgs 
(CLAH) model. 
When $\beta\to\infty$ the theory becomes a $Z_q$ gauge theory which 
is known to have a deconfining transition in $2+1$ dimensions. 
For $\kappa\to\infty$ 
we obtain the $XY$ model, which also undergoes a phase transition. 
The critical points corresponding to these two limiting cases 
are connected by a critical line separating the confining 
from the deconfining phase. There is no Coulomb phase 
in $2+1$ dimensions. The phase diagram and the critical properties 
of the action (\ref{AHM}) were studied in detail using 
large scale Monte Carlo simulations in Ref. \cite{Sudbo}. 
The CLAH model appears in many contexts in condensed matter 
physics. We shall cite only two important recent examples.  
In the $q=2$ version it arises as an effective theory 
for the two-dimensional quantum Heisenberg antiferromagnet (QHA)  
\cite{Sachdev}.  
There the gauge field originates from the Berry phase.  
Another recent example corresponds to the strongly correlated 
limit of a bosonic model considered by Senthil and Motrunich \cite{SM}. 
In this context the CLAH model describes a transition from an ordinary 
Mott insulating phase to a fractionalized insulating phase \cite{SM}.

However, bosonic actions like the one in Eq. (\ref{AHM}) are 
only part of the complete effective action associated with 
a doped Mott insulator. It usually contains also fermions coupled 
to a gauge field. \cite{IL,LN} The traditional route starts 
with a slave boson representation of the $t-J$ model where 
the projected electron operator is written as a 
composite particle, $c_{i\alpha}=b_i^\dagger f_{i\alpha}$, where 
$b_i$ is an auxiliary boson and $f_{i\alpha}$ is an auxiliary 
fermion. The auxiliary fields obey the constraint 
$b_i^\dagger b_i+ \sum_\alpha f_{i\alpha}^\dagger f_{i\alpha}=1$ 
at each lattice site. In the resulting effective 
action the auxiliary fields interact through a compact gauge field. 
The most popular way of doing things attribute the charge of 
electron to the bosonic field. Thus, the fermion would have no charge and 
would only carry the spin degree of freedom. If the effective model 
undergoes a deconfinement transition fermions and bosons will have 
an independent dynamics and in this way we can say the spin and 
charge are separated. Nayak \cite{Nayak} pointed out recently that things are 
not so simple because the assignment of the whole 
electron charge to the auxiliary boson is completely arbitrary. We 
could well say that the boson carries charge $\delta e$ while the 
fermion carries charge $(1-\delta)e$. The constraints of the 
theory, which are enforced by the coupling to the gauge field, would 
then ensure that the physics is unaffected by this arbitrary choice. 
This scenario implies that the auxiliary fields should not be 
associated with the physical charge and spin excitations of the 
theory. The auxiliary fields introduced this way are not part 
of the physical spectrum. This situation is reminiscent of the 
analysis made by Mudry and Fradkin some time ago. \cite{Mudry} 


In this paper, we review some recents results on the model 
(\ref{AHM}). We start discussing in the 
Section \ref{sec2} the finite-size scaling (FSS) analysis 
of the model in the $q>1$ case. We employ a new and very accurate 
method for extracting the critical exponents. In Section \ref{sec3} 
we discuss the phase diagram, which is obtained for  
$q=2,3,4,5$. In Section \ref{sec4} we discuss the $q=1$ case, 
whose phase diagram is presently a matter of intensive 
debate \cite{KNS,Herbut,Chernodub}. 

\section{The universality class of the deconfinement 
transition for $q>1$}
\label{sec2}    

In 1981 Bhanot and Freedman \cite{Bhanot} made a finite-size scaling 
(FSS) analysis of the CLAH model for $q=2$ and $d=2+1$. They obtained 
the phase diagram of the model but the critical exponents were not 
calculated. In order to determine the universality class of the model it 
is necessary to compute the critical exponents. The accurate determination 
of the critical behavior is not an easy task. In principle, on the basis 
of the results of Fradkin and Shenker \cite{FradShe}, we might think that 
the universality class will be the same as 
of the $Z_q$ spin model, except for the limit $\kappa\to\infty$ 
where the universality class is obviously that of the $XY$ model. 
If we assume that this is indeed so, we obtain that the critical 
exponents for the $q=2$ case have Ising values. However, we have 
recently shown through large scale Monte Carlo simulations 
that the situation is more complicated \cite{Sudbo}. For example, for $q=2$ 
we have obtained that the exponents are those of the Ising 
model in a large part of the phase diagram in the $\kappa\beta$-plane, 
but there is a finite interval $\kappa_1<\kappa<\kappa_2$ where 
the exponents vary continuously, before reaching $XY$ values 
for $\kappa>\kappa_2$. Therefore, it seems that the model 
features a fixed line rather than a fixed point and belongs 
to a new universality class.  

\begin{figure}
\label{m3}
\centerline{\scalebox{0.6}{\includegraphics{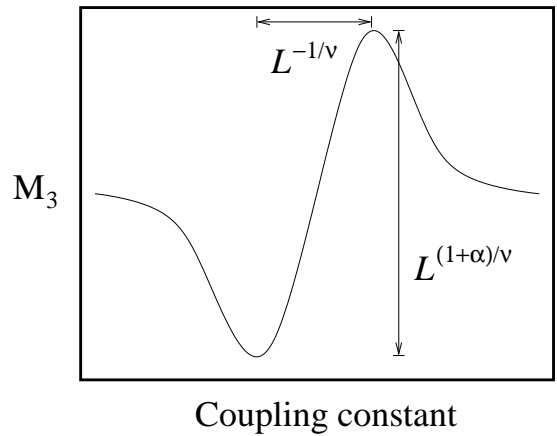}}}
\caption{Generic third moment $M_3$, showing how 
FSS is used to extract $\alpha$ and $\nu$}
\end{figure}

The FSS analysis of Bhanot and Freedman \cite{Bhanot} relies on the second 
moment of the free energy, i.e., the specific heat. In this case it is well 
known that very large system sizes are needed in order to 
identify a genuine critical behavior to high accuracy. We have 
shown that much better results are obtained by performing 
a FSS analysis based on the third moment of the free energy. Such a 
procedure is not only more accurate: it allows also for an 
independent determination of the exponents $\nu$ and $\alpha$, 
providing in this way a check of hyperscaling \cite{Sudbo}. The 
reason for this lies in the double-peak structure of the third moment. 
The FSS in this case is such that the width between the negative and 
positive peak scales as $L^{-1/\nu}$, while the height of the 
positive peak with respect to the negative one scales 
as $L^{(1+\alpha)/\nu}$. This scaling behavior is shown 
schematically for a generic third moment $M_3$ in Fig. \ref{m3}. 

\begin{figure}
\label{exps-q2}
\centerline{\scalebox{0.6}{\includegraphics{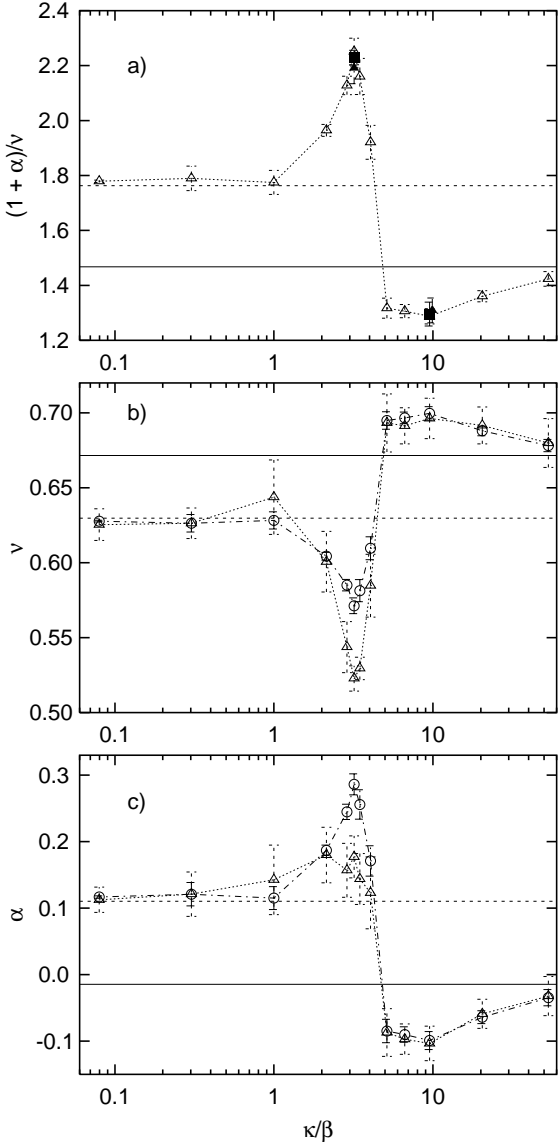}}}
\caption{a)  $(1+\alpha)/\nu$  from 
FSS finite-size of $M_3$ for Eq. (\ref{AHM}) for $q=2$. Note the
variation relative to the $Z_2$-limit $1.76$ (dotted horizontal line)  
and the $U(1)$-limit $1.467$ (solid horizontal line). b) Same for the 
exponent $\nu$, computed directly from $M_3$ ($\triangle$) and combining 
results for $(1+\alpha)/\nu$ with hyperscaling ($\bigcirc$). 
c) $\alpha$ as computed directly from $M_3$ 
($\triangle$) and using results for $(1+\alpha)/\nu$ with hyperscaling 
($\bigcirc$). The maximum and minimum in a) have been obtained by crossing 
the critical line along the trajectory $\beta(\kappa) = \beta_c + 
a ~ (\kappa - \kappa_c)$ with $a = \infty$ ($\triangle$), $a = 1$ 
($\blacksquare$), and $a=-1$ ($\blacktriangle$) using $\beta_c = 
0.665, \kappa_c = 2.125$ (max.), and $\beta_c = 0.525, \kappa_c = 
5.0$ (min.).}
\end{figure}

The critical exponents as a function of $\kappa/\beta$ are 
shown in Fig. \ref{exps-q2} for $q=2$. It is clearly seen 
that the Ising critical behavior connects to the $XY$ behavior 
through a regime of continuously varying critical exponents. 

\section{The phase diagram for $q>1$}
\label{sec3}

We have obtained the phase diagram for $q=2,3,4,5$. Among these 
values of $q$, only the $q=3$ case exhibits a first-order 
phase transition up to some point where it changes to second-order. 
This behavior is consistent with the fact that the three-state 
Potts model exhibits a first-order transition in three dimensions. 
This model is dual to the $Z_3$ gauge theory. The phase 
diagram for $q=3$ case therefore features a tricritical point. 
The tricritical point in the phase diagram is estimated as 
$(\beta_{\rm tri},\kappa_{\rm tri})=(1.23\pm0.03,1.73\pm0.03)$, 
corresponding to a ratio 
$\kappa_{\rm tri}/\beta_{\rm tri}=1.39\pm0.06$. At this point 
it is worth  making the following remark. The non-compact version of 
the model (\ref{AHM}) is known to exhibit a second-order phase 
transition for all values of $q$. As matter of fact, the value 
of $q$ is not important in that case, and can be absorbed 
into a redefinition of the gauge field. The universality class 
corresponds to the so called ``inverted'' $XY$ transition 
\cite{Dasgupta}. However, if the radial part of the scalar 
field is allowed to fluctuate, thus generalizing the non-compact 
version of (\ref{AHM}), then it is possible to show using 
duality arguments that the resulting model has a tricritical 
point approximately at $\kappa_{\rm tri}/\beta_{\rm tri}\approx 0.64$ 
\cite{Kleinert}. This estimate was confirmed recently by large 
scale Monte Carlo simulations \cite{Mo}. In the case of the 
$q=3$ CLAH model there is no need to consider the fluctuations of 
the radial component of the field to obtain a tricritical point. 
The origin of this tricritical point is completely different from 
the one of the non-compact model. Anyway, it is interesting to 
note that the ratio $\kappa_{\rm tri}/\beta_{\rm tri}$ in 
the $q=3$ CLAH model is more than twice larger than the corresponding 
ratio in the non-compact case.

In the context of the model discussed recently by Senthil and 
Motrunich \cite{SM}, the phase diagram for the $q=2$ case has 
the following physical meaning. The confined phase corresponds 
to gap to integer charge excitations and is therefore interpreted 
as an ordinary Mott insulating phase. The deconfined phase, on 
the other hand, is the same as the Higgs phase. This gives 
a gap to half-integer charge excitations and is interpreted 
as a {\it fractionalized} insulator. The model can in principle 
be artificially realized with present day technology 
by building an array of Josephson junctions of a particular 
type \cite{SM}.

\begin{figure}[htbp]
\centerline{\scalebox{0.35}{\rotatebox{270.0}{\includegraphics{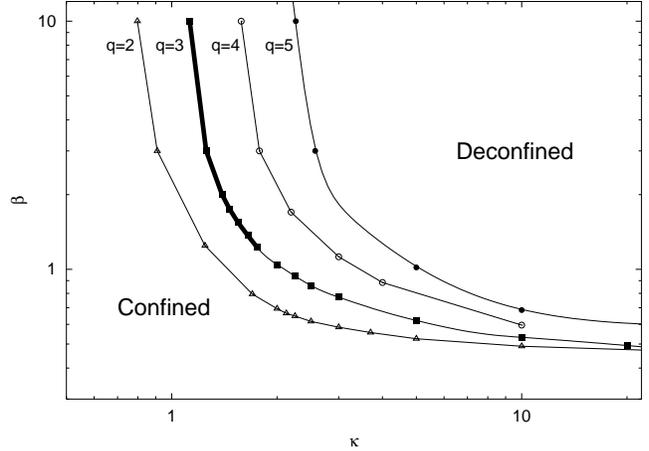}}}}
\caption{\label{phaseq=2} 
The phase diagram for the $d=3$ compact abelian Higgs model in three 
dimensions for $q=2,3,4,5$. All lines are critical  for all values of $\kappa$ 
except for the case $q=3$, which is first order for $\kappa < \kappa_{\rm{tri}}$
and second order otherwise. The thick solid portion of 
the $q=3$ line is a first-order transition line. 
The critical line approaches
the $3DXY$ value $\beta_c=0.453$ as $\kappa \to \infty$ for all integer values of 
$q>1$.}
\end{figure}

\section{The $q=1$ case}
\label{sec4}

The $q=1$ case was generally believed to not exhibit any phase transition 
at all. The coupling to matter fields would not change the scenario 
of the $d=2+1$ compact Maxwell theory, where it is known that 
no phase transition takes place \cite{Polyakov}: the electric charges 
would never deconfine. A research initiated with the study of 
the QHA showed that things seem not to be so simple \cite{IL}. 
The main point is that by integrating out the matter fields we 
obtain to quadratic order in the gauge field fluctuations the 
effective action

\begin{equation}
\label{anomMaxwell}
S_{\rm eff}\propto\int d^3xF_{\mu\nu}\frac{1}{\sqrt{
-\partial^2}}F_{\mu\nu}.
\end{equation}
Therefore, while in the absence of matter fields the monopoles 
interact through a $1/r$ potential, in the presence of matter 
fields this behavior changes to $-\ln r$. In the former case 
it is well known that upon dualizing the theory a  
sine-Gordon action is obtained. In the latter case, however, an 
{\it anomalous} sine-Gordon action arises \cite{KNS}:

\begin{equation}
\label{ASG}
S=\frac{1}{8\pi^2K}
\int d^3 x[\varphi(-\partial^2)^{3/2}\varphi
-2z\cos\varphi], 
\end{equation}
where $K=1/g^2$, with $g$ being the gauge coupling. Thus, we 
arrive at a behavior similar to the two-dimensional case. 
By simply looking at the scaling behavior of the fugacity of 
the monopole gas, one would be lead to conclude that a 
Kosterlitz-Thouless (KT) phase transition happens in three 
space-time dimensions. 
{\it Note that this KT transition is at zero temperature and has 
nothing to do with other KT transitions obtained by 
dimensional reduction due to temperature effects}. However, 
it is not enough to study the scaling of just the fugacity. Since 
we are in $d=2+1$ a more thorough analysis is needed in 
order to fully establish the KT behavior. To this end, it is 
necessary to also know the scaling behavior of the ``stiffness'' 
$K$, since the scaling of the fugacity and the scaling of the 
stiffness mutually influence each other in a subtle way. 

For the usual sine-Gordon model in $d$ dimensions we obtain the 
following coupled recursion relations for the fugacity and the
stiffness \cite{Kosterlitz}

\begin{equation}
\label{flowK}
\frac{d K^{-1}}{dl}=4\pi^2 y^2-(2-d)K^{-1},
\end{equation}

\begin{equation}
\label{flowy}
\frac{dy}{dl}=\left[d-2\pi^2f(d)K\right]y,
\end{equation}
where $f(d)=(d-2)\Gamma[(d-2)/2]/(4\pi)^{d/2}$. For $d=2$ 
the above equations imply the existence of a line of fixed points 
characteristic of the KT transition. For $d=3$, however, 
no fixed point is found and therefore no phase transition 
takes place. This is consistent with Polyakov's result for 
the compact Maxwell theory in $d=2+1$ \cite{Polyakov}. The issue 
here is how the above recursion relations are modified in the case of 
the anomalous sine-Gordon model, Eq. (\ref{ASG}). To investigate this 
in great generality we will consider screening 
in a Coulomb gas of monopoles with a propagator 
of the form $1/|p|^\sigma$ in $2\leq d<4$ dimensions with 
$\sigma>0$. The particular case $d=\sigma=3$ corresponds to our 
anomalous sine-Gordon theory. The bare potential is given by 
$U_0(r)=-4\pi^2KV(r)$, where 

\begin{equation}
\label{potential}
V(r)=\frac{\Gamma\left(\frac{d-\sigma}{2}\right)}{
2^\sigma\pi^{d/2}\Gamma(\sigma/2)}[(\Lambda r)^{\sigma-d}-1],
\end{equation}
with $\Lambda$ being 
an ultraviolet cutoff. Note that for $d=\sigma$ we obtain a 
potential $\propto\ln r$. The bare electric field is given by 
$E_0(r)=-4\pi^2K A(d,\sigma)r^{\sigma-d-1}/r_0^{\sigma-d}$, where 
$r_0\equiv 1/\Lambda$ and 
$A(d,\sigma)=(d-\sigma)\Gamma[(d-\sigma)/2]/[2^\sigma\pi^{d/2}\Gamma
(\sigma/2)]$. The  {\it bare} electric field is renormalized by the other 
dipoles which are treated as a dielectric medium. The {\it renormalized} 
electric field is then given by

\begin{equation} 
E(r)=-\frac{4\pi^2K  ~ A(d,\sigma) ~r^{\sigma-d-1}}{\varepsilon(r)}, 
\end{equation}
where $\varepsilon(r)$ is the scale-dependent dielectric constant of the 
medium. This problem can be solved self-consistently for 
a renormalized potential $U(r)$ derived from the above 
electric field \cite{KNS}. In this way we obtain the 
recursion relations 

\begin{equation}
\label{flowK2}
\frac{d K^{-1}}{dl}=4\pi^2y^2-(\sigma-d)K^{-1},
\end{equation}

\begin{equation}
\label{flowy2}
\frac{dy}{dl}=\left[d-\eta_y-2\pi^2A(d,\sigma)K\right]y,
\end{equation}
where the {\it anomalous dimension} of the fugacity is given by 
$\eta_y=(\sigma-2)/2$. The above equations reduce to Eqs. (\ref{flowK}) 
and (\ref{flowy}) when $\sigma=2$. The case relevant to the 
$q=1$ CLAH model is $\sigma=d=3$. In this case 
Eqs. (\ref{flowK2}) and (\ref{flowy2}) are very similar to the usual KT 
recursion relations, except for the presence of the anomalous scaling 
dimension of the fugacity, $\eta_y$ which is nonzero in our case and 
given by $\eta_y=1/2$. Thus, we obtain that the monopole gas undergoes 
a KT-like phase transition. This is in contrast with Polyakov's compact 
Maxwell theory where the monopoles are always in the plasma 
phase. Using the usual duality arguments, we identify the 
plasma phase with the confined phase for the electric charges. 
The dielectric phase of the monopole gas is accordingly identified 
with the deconfined phase for the electric charges.  

At this point an important remark is in order. The above screening 
analysis was made in real space, in the spirit of the original 
Kosterlitz and Thouless paper \cite{KT}. Indeed, we have considered 
a {\it local} scale dependent dielectric constant. Screening 
arguments in momentum space usually involve a dielectric 
constant which is local in momentum space, leading to 
an electric displacement vector of the form 
${\bf D}({\bf r})=\int d^d r' \epsilon({\bf r}-{\bf r}'){\bf E}({\bf r}')$, 
while in our case we have simply 
${\bf D}({\bf r})=\epsilon({\bf r}){\bf E}({\bf r})$. 
In the present context this may be understood by using a 
classical electrostatic argument. Let us consider the potential 
(\ref{potential}) in the case $\sigma=d$, such that we have 
a $\ln r$-potential in $d$ dimensions. Due to Gauss's theorem 
in $d$-dimensions, the field equation is given by \cite{Dixit} 
$\nabla\cdot({\bf E} r^{2-d})=S_d\rho(r)$, where $S_d$ is the 
solid angle in $d$-dimensions. This can be cast in a more familiar 
form $\nabla\cdot{\bf D}=S_d r_0^{d-2}\rho(r)$ by introducing  
the ``dielectric constant'' $\varepsilon_0(r)=(r/r_0)^{2-d}$. 
Note that for $d>2$ this dielectric constant vanishes at large 
distances. This is rather unusual since in classical electromagnetism 
it can be shown that the dielectric constant is always greater 
than one. However, such a situation corresponds precisely to the 
anti-screening mechanism discussed many years ago in theories of 
confinement \cite{Kogut}. Thus, already at this level, even before 
taking the dipoles into account, we can define a kind of scale-dependent 
dielectric constant which is local in real space.

\section{Conclusions}

In this paper we discussed the deconfinement transition in 
the CLAH model, Eq. (\ref{AHM}), in $d=2+1$. To each value of 
the charge $q$ corresponds a distinct universality class, in 
contrast to the non-compact model. While a deconfined 
transition is expected for $q>1$, the case $q=1$ is still 
controversial \cite{KNS,Herbut,Chernodub} and needs further study.   
One feature that also deserves further study is the possible 
existence of a fixed line in the $q=2$ case. The numerical 
results strongly suggest this possibility but multicritical 
fixed points are also possible. 

\begin{acknowledgement}
FSN would like to thank Boris Pioline and Christoph Berger 
for the invitation to give the talk in the EPS HEP2003 
conference, upon which this paper is based. We thank the 
Deutsche Forschungsgemeinschaft (DFG) and the Norwegian 
Research Council, Project No. 131520/432, for  financial 
support.
\end{acknowledgement}   

%
%
%
%
%

\end{document}